\RequirePackage{ifpdf}
\documentclass[11pt,a4paper]{article}
\pdfoutput=1 
\usepackage{amsmath} 
\usepackage{jheppub}
\usepackage{pstricks}
\usepackage{longtable}
\usepackage[final]{pdfpages}
\usepackage{ifpdf}
\usepackage{flexisym}
\usepackage{breqn}
\usepackage{slashed}
\usepackage{graphicx}
\usepackage{caption}
\usepackage{subcaption}

\def\ep{\epsilon}
\def\unM{\hat{\cal M}}

\def\rnM{{\cal M}}

\def\b0{\beta_0}
\def\cD{{\cal D}}

\title{Spin-2 Form Factors at Three Loop in QCD}

\author[a]{Taushif Ahmed,}
\author[b]{Goutam Das,}
\author[b]{Prakash Mathews,}
\author[a]{Narayan Rana}
\author[a]{and V. Ravindran}

\affiliation[a]{The Institute of Mathematical Sciences,\\ IV Cross Road, CIT Campus, Chennai 600 113, India}
\affiliation[b]{Saha Institute of Nuclear Physics,\\ 1/AF Bidhan Nagar, Kolkata 700 064, India}

\emailAdd{taushif@imsc.res.in}
\emailAdd{goutam.das@saha.ac.in}
\emailAdd{prakash.mathews@saha.ac.in}
\emailAdd{rana@imsc.res.in}
\emailAdd{ravindra@imsc.res.in}

\abstract{
Spin-2 fields are often candidates in physics beyond the Standard Model namely the models with extra-dimensions
where spin-2 Kaluza-Klein gravitons couple to the fields of the SM.  Also, in the context
of Higgs searches, spin-2 fields have been studied as an alternative to the scalar Higgs boson. 
In this article, we present the complete three loop QCD radiative corrections to the spin-2
quark-antiquark and spin-2 gluon-gluon form factors in SU(N) gauge
theory with $n_f$ light flavors. These form factors contribute
to both quark-antiquark and gluon-gluon initiated processes
involving spin-2 particle in the hadronic reactions at the LHC.   
We have studied the structure of infrared singularities in these form factors up to three loop level
using Sudakov integro-differential equation and found
that the anomalous dimensions originating from soft and collinear regions of the
loop integrals coincide with those of the electroweak vector boson and Higgs form factors
confirming the universality of the infrared singularities in QCD amplitudes.  
} 

\begin{document}
\allowdisplaybreaks[4]
\unitlength1cm
\maketitle
\flushbottom

\vspace{1cm}
\section{Introduction}
\setcounter{equation}{0}
\label{sec:intro}

In the context of the recent discovery of the new boson at the LHC, with
mass of about 125 GeV \cite{Aad:2012tfa, Chatrchyan:2012xdj}, there has
been renewed interest in massive spin-2 resonance which could also lead
to similar final states \cite{Ellis:2012mj}.  The massive spin-2 could be
a Kaluza-Klein (KK) graviton of the TeV scale gravity models 
\cite{ADD, Randall:1999ee} as a result of gravity propagating in the extra dimensional 
bulk
or any generic spin-2 resonance in some other new physics scenarios.  It
was noted in \cite{Fok:2012zk} that gauge symmetry and Lorentz invariance
forbid operators of dimension four that could lead to a coupling of a
massive spin-2 resonance to a pair of SM particles.  Further, if the
flavor and CP symmetries of the SM are respected by these new physics
scenarios, the leading dimension five operator is none other than the
energy momentum tensor $T_{\mu \nu}$ of the SM particles.  The structure
of the operator coupling thus being identical to the KK graviton, though
the constant coefficients could be different for the KK graviton or any
generic spin-2 imposter.  Nonetheless, methods to distinguish KK graviton
from the imposter have been proposed \cite{Fok:2012zk} and will be of
importance for BSM searches at the LHC which is now operational at higher
energies and luminosity.  The increasing accuracy of the experimental data
at the LHC Run-II, demands an equally precise theoretical predictions.

To match the current theoretical accuracy of say 
the Drell-Yan production \cite{Hamberg:1990np, Ahmed:2014cla, Catani:2014uta},
the Higgs boson production in gluon fusion \cite{Harlander:2002wh, Anastasiou:2002yz, Ravindran:2003um, Anastasiou:2014vaa, Li:2014bfa, deFlorian:2014vta, Anastasiou:2015ema}, in 
bottom quark annihilation \cite{Harlander:2003ai, Ahmed:2014cha} and associated production
with a vector boson \cite{Brein:2003wg, Kumar:2014uwa}
at the LHC, it is imperative that competing BSM models are
also available to the same accuracy in higher orders in QCD. 
Form factors are essential ingredients for many precision calculation
in QCD.  An important building block for phenomenological study is the
computation of $q \bar q \to$ spin-2 and $g g \to$ spin-2 form factors
and is at present 
available to two-loop in QCD \cite{deFlorian:2013sza}, while for many
processes of interest they are now available to the 3-loop
order \cite{Moch:2005tm, Moch:2005id, Baikov:2009bg, Gehrmann:2010ue, Gehrmann:2010tu, Gehrmann:2014vha}.
Stringent bounds \cite{Aad:2014cka, Chatrchyan:2012oaa, Aad:2015mna} on the parameters of ADD and RS models are available due to the
presence of precise theoretical predictions for various
important observables up to NLO level in QCD.  Often, these observables
suffer from large uncertainties resulting from renormalization and factorization
scales and the only remedy to this is to include higher order QCD effects
to the born contributions.  The NLO QCD predictions based on fixed order as well as
parton shower improved in the M{\sc ad}G{\sc raph}5\_{\sc a}MC@NLO \cite{Alwall:2014hca}  framework for
di-final state \cite{Mathews:2004xp, Mathews:2005zs, Kumar:2006id, Frederix:2013lga, Kumar:2009nn, Frederix:2012dp, Agarwal:2009zg, Agarwal:2010sp, Das:2014tva}
productions in the gravity mediated models have already played important role in constraining the
model parameters of ADD and RS. 
Next-to-next-to leading order (NNLO) corrections for the graviton production at the LHC in the threshold limit are already available \cite{deFlorian:2013wpa} and
attempts to improve these predictions through
NNLO corrections and beyond are already underway \cite{Ahmed:2014gla}.  As these corrections are
only sensitive to the tensorial interaction and not sensitive to the details of the model,
these results are applicable to production of any generic spin-2 resonance.  Hence this
article takes the first step towards going beyond NNLO for the resonant production
of a generic spin-2 particle at the LHC, namely the computation of quark and
gluon form factors at three loop level in perturbative QCD with $n_f$ light flavors.
We will report the first results on the threshold effects at N$^3$LO in the future publication
and demonstrate the importance such corrections at the LHC in the context of spin-2 resonance
searches.

In addition to the phenomenological importance with respect
to precise predictions of some observable, form factors in QCD
are of considerable theoretical interest in terms of the
factorization and universal nature of the singular structure.  
Studying the infrared pole structure and factorization properties 
of these IR singularities in multi-loop QCD amplitudes with tensorial
coupling to 3-loop order and to confirm the standard expectation of 
QCD amplitudes 
\cite{Catani:1998bh,Sterman:2002qn, Becher:2009cu,Gardi:2009qi}
is an essential prerequisite.  
The spin-2 field being a tensor of rank-2 is coupled to the energy-momentum
tensor $T_{\mu \nu}$, which is a symmetric and conserved quantity.
The operator $T_{\mu \nu}$ of QCD is finite \cite{Nielsen:1977sy}, which
would imply no UV renormalization is required.  Further $T_{\mu \nu}$
consists of gauge invariant terms and in addition gauge dependent and ghost terms,
we explicitly observe that to the 3-loop order these form factors are 
independent of the gauge dependent and ghost terms 
\cite{Nielsen:1977sy,Zoller:2012qv}, which is an important check of the calculation.
From a computational point of view 3-loop amplitudes with higher tensorial
coupling is being attempted for the first time.  At the intermediate 
stages of the computation this leads to higher rank tensorial integrals resulting
from more than 3000 three loop Feynman amplitudes contributing to the gluon form factor alone.
This computation again establishes the power of several state-of-the-art techniques namely IBP and LI identities
and differential equation method to solve the master integrals.

In the next section \ref{sec:lag}, we describe the effective Lagrangian. In section \ref{sec:ff}, after defining the quark and 
gluon form factors, we present the computational details at three loop level followed by the results. The details of ultraviolet
renormalization and universal structure of infrared poles are given in section \ref{sec:uv} and section \ref{sec:ir} respectively.
Finally we conclude with our findings in section \ref{sec:conclu}.


\section{The Effective Lagrangian}
\label{sec:lag}
The effective Lagrangian that describes the interaction of the spin-2 field with the SM fields can
be written down in a gauge invariant way through the energy momentum tensor of the SM fields.
We denote the spin-2 field by $h^{\mu\nu}$ and  
the SM energy momentum tensor by $T_{\mu\nu}^{SM}$.  Since we are interested only in the
QCD corrections to processes involving spin-2 fields, 
we restrict ourselves to the QCD part of $T_{\mu\nu}^{SM}$ and the corresponding  
action reads \cite{ADD, Randall:1999ee} as
\begin{eqnarray}\label{intlag}
{\cal S} = {\cal S}_{SM} +{\cal S}_h -  \frac{\kappa}{2} \int d^4 x ~T^{QCD}_{\mu\nu}
(x)~ h^{\mu\nu} (x) \, ,
\end{eqnarray}
where ${\cal S}_{SM}$ is the SM action, ${\cal S}_{h}$ is the kinetic energy part of the action corresponding to spin-2 fields, $\kappa$ is a dimensionful coupling and $T^{QCD}_{\mu\nu}$ is the energy momentum tensor of QCD given by
\begin{eqnarray}\label{emT}
T^{QCD}_{\mu\nu} &=& -g_{\mu\nu} {\cal L}_{QCD} - F_{\mu\rho}^a F^{a\rho}_\nu
- \frac{1}{\xi} g_{\mu\nu} \partial^\rho(A_\rho^a\partial^\sigma A_\sigma^a)
+ \frac{1}{\xi}(A_\nu^a \partial_\mu(\partial^\sigma A_\sigma^a) + A_\mu^a\partial_\nu
(\partial^\sigma A_\sigma^a))
\nonumber\\[1ex]
&&+\frac{i}{4} \Big[ \overline \psi \gamma_\mu (\overrightarrow{\partial}_\nu -i g_s T^a A^a_\nu)\psi
-\overline \psi (\overleftarrow{\partial}_\nu + i g_s T^a A^a_\nu) \gamma_\mu \psi
+\overline \psi \gamma_\nu (\overrightarrow{\partial}_\mu -i g_s T^a A^a_\mu)\psi
\nonumber\\[1ex]
&&-\overline \psi (\overleftarrow{\partial}_\mu + i g_s T^a A^a_\mu) \gamma_\nu \psi\Big]
+\partial_\mu \overline \omega^a (\partial_\nu \omega^a - g_s f^{abc} A_\nu^c \omega^b)
\nonumber\\[1ex]
&&+\partial_\nu \overline \omega^a (\partial_\mu \omega^a- g_s f^{abc} A_\mu^c \omega^b).
\end{eqnarray}
$g_s$ is the strong coupling constant and $\xi$ is the gauge fixing parameter.  The $T^a$ are generators and $f^{abc}$
are the structure constants of $SU(3)$.   
Note that spin-2 fields couple to ghost fields ($\omega^a$) \cite{Mathews:2004pi} as well in order to cancel unphysical
degrees of freedom of gluon fields ($A^a_{\mu}$).

\section{The Form Factors}
\label{sec:ff}

The form factor parametrizes the interaction of the spin-2 field with those of SM order by order 
in perturbation theory.  We compute both quark and gluon form factors 
by sandwiching the energy-momentum tensor between
on-shell quark and gluon states respectively normalized by their respective born amplitudes:  

\begin{eqnarray}
\label{eq:miexp}
 \hat{\cal F}^{T}_{\rm I}(Q^2,\epsilon) &=& \frac{\unM_{\rm I}^{(0)*} . M_{\rm I}}{\unM_{\rm I}^{(0)*} . \unM_{\rm I}^{(0)}}
\nonumber\\
 &=& \sum_{n=0}^{\infty} \hat a_s^n \left( \frac{Q^2}{\mu^2} \right)^{n {\ep\over 2}} S_{\ep}^{n}  \hat{\cal F}^{T,(n)}_{\rm I}(\epsilon),
 \quad \quad \quad I=q,g
\end{eqnarray}
where $M_{\rm I}$ are the unrenormalized amplitudes computed in powers of the bare
strong coupling constant $\hat{a}_s = \hat{g}_s^2/16 \pi^2$ using dimensional regularization 
in $d=4+\epsilon$ dimensions, that is 
\begin{equation}
 M_{\rm I}(Q^2,\epsilon) 
 = \sum_{n=0}^{\infty} \hat a_s^n \left( \frac{Q^2}{\mu^2} \right)^{n {\ep\over 2}} S_{\ep}^{n}  
\hat{\cal M}^{(n)}_{\rm I}(\epsilon),
\end{equation}
where $Q^2 = -2 p_1\cdot p_2$ and $p_1$, $p_2$ are the momenta of external quark or gluon on-shell states. 
The dimensionful scale $\mu$ is introduced to keep the strong 
coupling constant dimensionless in $d$ space-time 
dimensions.  The other constant at $n^{\rm th}$ loop is    
$S_{\ep}^n = \exp[\frac{n \ep}{2} (\gamma_E - \ln 4\pi)]$ where Euler constant $\gamma_E = 0.5772 \ldots$. 

In \cite{deFlorian:2013sza}, both one and two loop form factors were presented in dimensional regularization and later
on, they were used in \cite{deFlorian:2013wpa} to compute the threshold corrections to 
Drell-Yan production at the LHC in ADD and RS models
to second order in strong coupling constant.  In the following, we present the third order
correction to the form factors in QCD.

\subsection{Computational Procedure}
In this section, we describe in detail the method that we follow to compute 
both quark and gluon form factors
of the energy momentum tensor to third order in strong coupling constant using dimensional 
regularization.  The relevant amplitudes are 
generated using QGRAF \cite{Nogueira:1991ex}.
At third order alone, there are 3374 and 1072 number of Feynman diagrams for gluon and quark form factors respectively.
The QGRAF generated amplitudes are then converted into a suitable 
format using routines developed using the symbolic manipulation program FORM \cite{Vermaseren:2000nd}. 
Both group as well as Lorentz indices are carefully handled to express the form 
factors in a suitable color basis involving Casimir operators of $SU(N)$ 
with the coefficients containing three loop scalar integrals.  For the gluon form factor
we have summed only the physical polarizations of the external gluons using 
\begin{equation}
 \sum_s \varepsilon^{\mu}(p_i,s) \varepsilon^{\nu *}(p_i,s) = -g^{\mu\nu} + \frac{p_i^{\mu} q_i^{\nu} + q_i^{\mu} p_i^{\nu}}{p_i.q_i}
\end{equation}
where, $p_i$ is the $i^{th}$-gluon momentum and $q_i$ is the corresponding light-like momentum.  We choose $q_1 = p_2$ and $q_2 = p_1$ for simplicity.  
For the external spin-2 fields, we have used the     
$d$ dimensional polarization sum given in \cite{Mathews:2004xp} with $q$ being the spin-2 momentum
\begin{eqnarray}
 B^{\mu \nu ; \rho \sigma}(q) &=& \left( g^{\mu\rho} - \frac{\, \, q^{\mu} q^{\rho} }{q.q} \right)  \left(g^{\nu\sigma} - \frac{\, \,q^{\nu} q^{\sigma} }{q.q} \right) 
                           + \left( g^{\mu\sigma} - \frac{\, \,q^{\mu} q^{\sigma} }{q.q} \right)  \left(g^{\nu\rho} - \frac{\, \,q^{\nu} q^{\rho} }{q.q} \right) 
\nonumber\\[1ex]
&& -\, \, \frac{2}{d-1} \left( g^{\mu\nu} - \frac{\, \,q^{\mu} q^{\nu} }{q.q} \right) \left( g^{\rho\sigma} - \frac{\, \,q^{\rho} q^{\sigma} }{q.q} \right) \,.
\end{eqnarray}
We have used Feynman gauge throughout.  

At three loop level, we find that the diagrams contributing to
form factors can have at most 9 independent propagators involving two external momenta $p_1,p_2$ and
three internal loop momenta $k_1,k_2,k_3$, while the maximum number of
scalar products that can appear in the numerator of each diagram can be 12.  
Hence we need to  
increase the number of propagators to 12 which allow us to
classify all the three loop diagrams into three different auxiliary topologies.  
We take the help of Reduze2 \cite{vonManteuffel:2012np} for this purpose.
The topologies \cite{Gehrmann:2010ue} that are used in our computation are given below 
\begin{align}
{\rm A}_1 &:~~ \cD_1, \cD_2, \cD_3, \cD_{12}, \cD_{13}, \cD_{23}, \cD_{1;1}, \cD_{1:12}, \cD_{2;1}, \cD_{2:12}, \cD_{3;1}, \cD_{3:12} 
\nonumber\\
{\rm A}_2 &:~~ \cD_1, \cD_2, \cD_3, \cD_{12}, \cD_{13}, \cD_{23}, \cD_{13;2}, \cD_{1:12}, \cD_{2;1}, \cD_{12:2}, \cD_{3;1}, \cD_{3:12} 
\nonumber\\
{\rm A}_3 &:~~ \cD_1, \cD_2, \cD_3, \cD_{12}, \cD_{13}, \cD_{123}, \cD_{1;1}, \cD_{1:12}, \cD_{2;1}, \cD_{2:12}, \cD_{3;1}, \cD_{3:12} 
\end{align}
where,
\begin{align}
\cD_{i} = k_{i}^2,~ \cD_{ij} = (k_i-k_j)^2,~ \cD_{ijl} = (k_i-k_j-k_l)^2,  \hspace{1cm}
\nonumber\\
\cD_{i;j} = (k_i-p_j)^2,~ \cD_{i;jl} = (k_i-p_j-p_l)^2,~ \cD_{ij;l} = (k_i-k_j-p_l)^2 \,.
\end{align}
%
%
The resulting integrals classified in terms of three topologies, are then reduced to a set of
master integrals by using a systematic approach that uses 
Integration by parts (IBP) \cite{chet} and Lorentz invariant (LI) \cite{gr} identities. 
The IBP identities follow from the fact that within dimensional regularization,
the integrals are finite and well-behaved and hence any integrand at the boundary must be zero.
Following this, the generalization of Gauss theorem implies
the integral of the total derivative with respect to any loop momenta to be zero, that is
\begin{equation}
 \int \frac{d^d k_1}{(2 \pi)^d} \cdot \cdot \cdot \int \frac{d^d k_3}{(2 \pi)^d} \frac{\partial}{\partial k_i} \cdot \left( v_j \frac{1}{\prod_l D_l^{n_l} } \right) = 0 \,,
\end{equation}
where $n_l$ is an element of $\vec n = (n_1,\cdot \cdot \cdot,n_{12})$ with $n_l \in Z$ and $D_l$s are propagators which depend on the loop and external momenta.
The four vector $v_j^\mu$ can be both loop and external momenta.
Performing the differentiation on the left hand side and expressing the scalar products of $k_i$ and $p_j$ linearly in terms of
$\cD_l$'s, one obtains the IBP identities given by
\begin{equation}
\sum_i a_i J(b_{i,1} + n_1, ... , b_{i,12} + n_{12} ) = 0 
\end{equation}
where
\begin{equation}
 J(\vec m) = J(m_1,\cdot \cdot \cdot ,m_{12})=
 \int \frac{d^d k_1}{(2 \pi)^d} \cdot \cdot \cdot \frac{d^d k_3}{(2 \pi)^d} \frac{1}{\prod_l D_l^{m_l}}   
\end{equation}
with $b_{i,j} \in \{-1,0,1\}$ and $a_i$ are polynomial in $n_j$.
The LI identities follow from the fact that the loop integrals are invariant under Lorentz transformations of the external momenta, that is
\begin{equation}
 p_i^{\mu} p_j^{\nu} \left(  \sum_k p_{k [\nu} \frac{\partial}{\partial p_k^{\mu]}}  \right) J(\vec n) = 0.
\end{equation}
For the case of three loop form factor, there are 15 IBP identities and 1 LI identity for each integrand, and hence there are large number 
of equations for the whole system. 
These equations can be solved to relate the large number of scalar integrals and express them in terms of a set of fewer integrals 
which are the so called master integrals. 
To solve this large system of equations, there are dedicated computer algebra tools like AIR \cite{Anastasiou:2004vj}, FIRE \cite{Smirnov:2008iw}, REDUZE \cite{Studerus:2009ye, vonManteuffel:2012np}, LiteRed \cite{Lee:2012cn, Lee:2013mka} etc.
We use the Mathematica based package LiteRedV1.82 along with MintV1.1 \cite{Lee:2013hzt}.

\begin{figure}[h!]
    \centering
    \begin{subfigure}[b]{0.23\textwidth}
        \includegraphics[width=\textwidth]{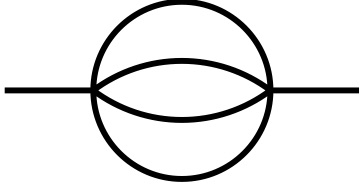}
        \caption*{$B_{4,1} \equiv A_{1,001101010000} $}
    \end{subfigure}
    ~ 
    \begin{subfigure}[b]{0.35\textwidth}
        \includegraphics[width=\textwidth]{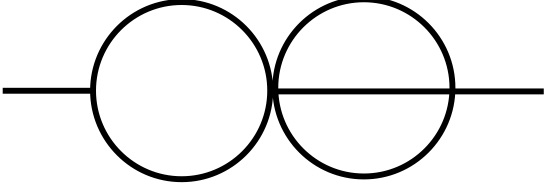}
        \caption*{$B_{5,1} \equiv A_{1,011010010100}$}
    \end{subfigure}
    ~ 
    \begin{subfigure}[b]{0.23\textwidth}
        \includegraphics[width=\textwidth]{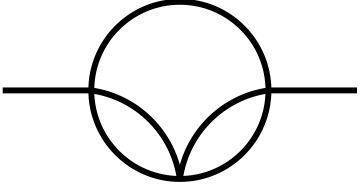}
        \caption*{$B_{5,2} \equiv A_{1,001011010100}$}
    \end{subfigure}

\vspace{0.5cm}

    \begin{subfigure}[b]{0.43\textwidth}
        \includegraphics[width=\textwidth]{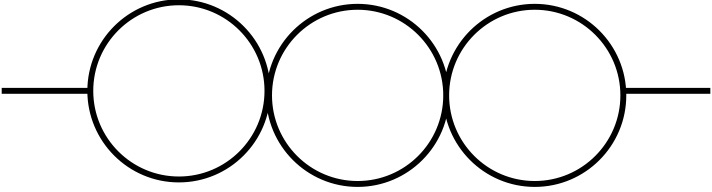}
        \caption*{$B_{6,1} \equiv A_{1,111000010101}$}
    \end{subfigure}
    ~ 
    \begin{subfigure}[b]{0.23\textwidth}
        \includegraphics[width=\textwidth]{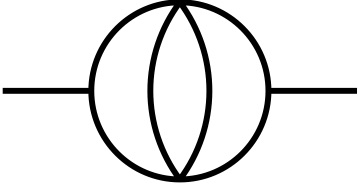}
        \caption*{$B_{6,2} \equiv A_{1,011110000101}$}
    \end{subfigure}
    ~~ 
    \begin{subfigure}[b]{0.23\textwidth}
        \includegraphics[width=\textwidth]{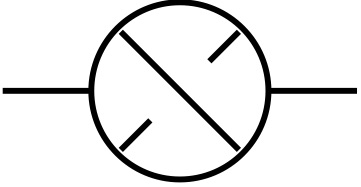}
        \caption*{$B_{8,1} \equiv A_{3,011111010101}$}
    \end{subfigure}

\vspace{0.5cm}

    \begin{subfigure}[b]{0.31\textwidth}
        \includegraphics[width=\textwidth]{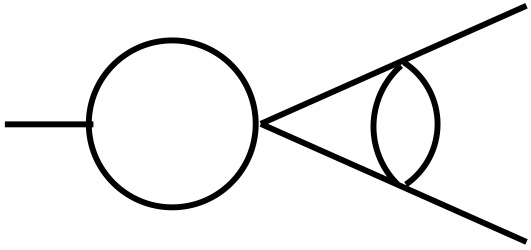}
        \caption*{$C_{6,1} \equiv A_{1,011100100101}$}
    \end{subfigure}
    ~  
    \begin{subfigure}[b]{0.31\textwidth}
        \includegraphics[width=\textwidth]{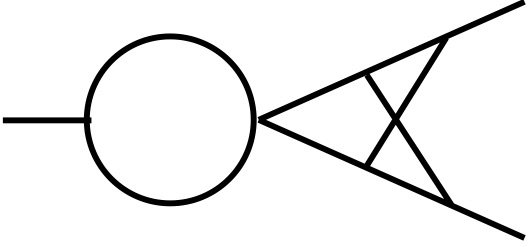}
        \caption*{$C_{8,1} \equiv A_{2,111100011101}$}
    \end{subfigure}
    \caption{Two point and factorizable three point three loop integrals}
    \label{fig:BnC}
\end{figure}

\begin{figure}[h!]
    \centering
    \begin{subfigure}[b]{0.22\textwidth}
        \includegraphics[width=\textwidth]{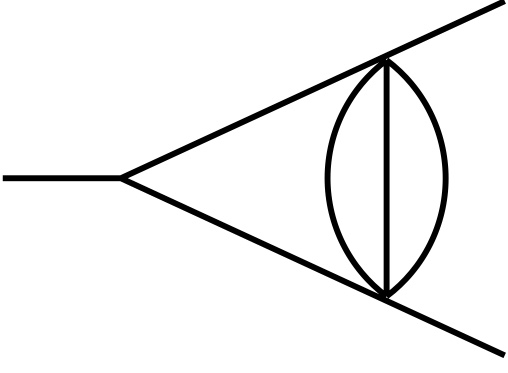}
        \caption*{$A_{5,1} \equiv A_{1,001101100001}$}
    \end{subfigure}
    ~~  
    \begin{subfigure}[b]{0.22\textwidth}
        \includegraphics[width=\textwidth]{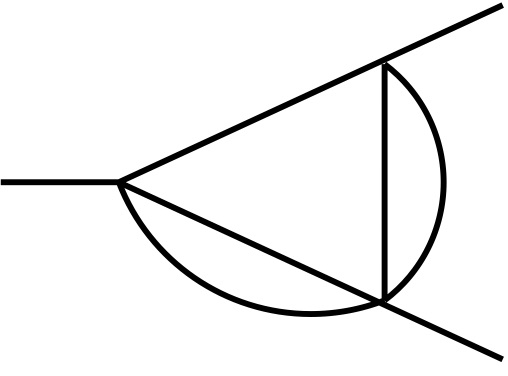}
        \caption*{$A_{5,2} \equiv A_{1,001011011000}$}
    \end{subfigure}
    ~~ 
    \begin{subfigure}[b]{0.22\textwidth}
        \includegraphics[width=\textwidth]{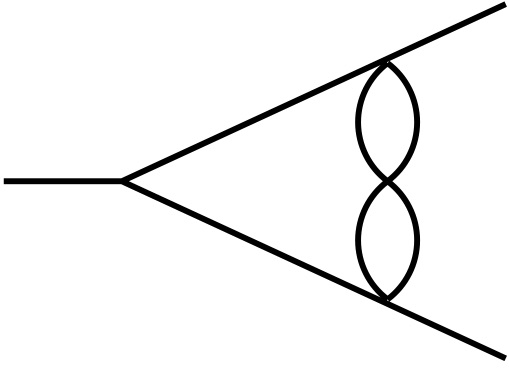}
        \caption*{$A_{6,1} \equiv A_{1,010101100110}$}
    \end{subfigure}

\vspace{0.5cm}

    \begin{subfigure}[b]{0.22\textwidth}
        \includegraphics[width=\textwidth]{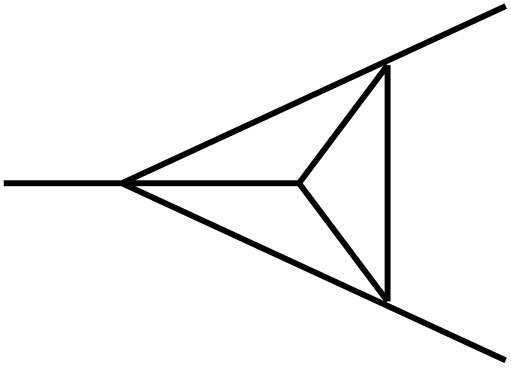}
        \caption*{$A_{6,2} \equiv A_{1,001111011000}$}
    \end{subfigure}
    ~~ 
    \begin{subfigure}[b]{0.22\textwidth}
        \includegraphics[width=\textwidth]{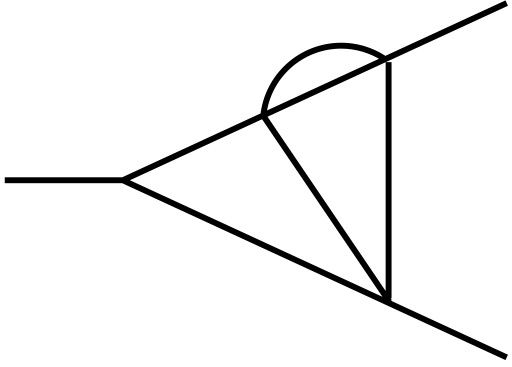}
        \caption*{$A_{6,3} \equiv A_{1,001110100101}$}
    \end{subfigure}
    ~~  
    \begin{subfigure}[b]{0.22\textwidth}
        \includegraphics[width=\textwidth]{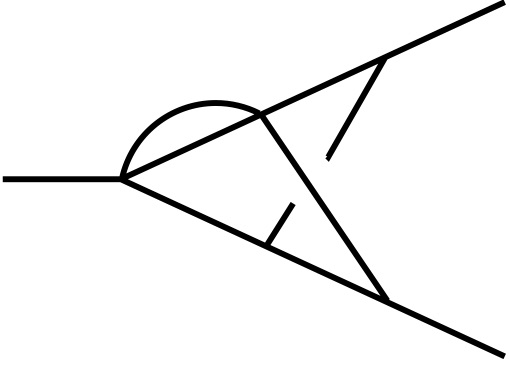}
        \caption*{$A_{7,1} \equiv A_{2,011110011100}$}
    \end{subfigure}

\vspace{0.5cm}

    \begin{subfigure}[b]{0.22\textwidth}
        \includegraphics[width=\textwidth]{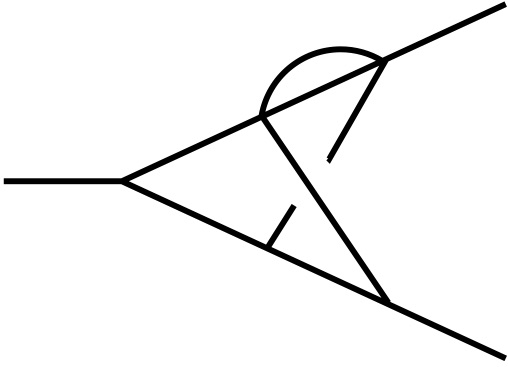}
        \caption*{$A_{7,2} \equiv A_{2,011011001101}$}
    \end{subfigure}
    ~~ 
    \begin{subfigure}[b]{0.22\textwidth}
        \includegraphics[width=\textwidth]{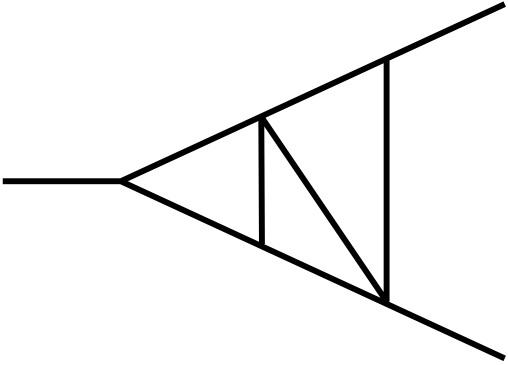}
        \caption*{$A_{7,3} \equiv A_{1,011011110100}$}
    \end{subfigure}
    ~~  
    \begin{subfigure}[b]{0.22\textwidth}
        \includegraphics[width=\textwidth]{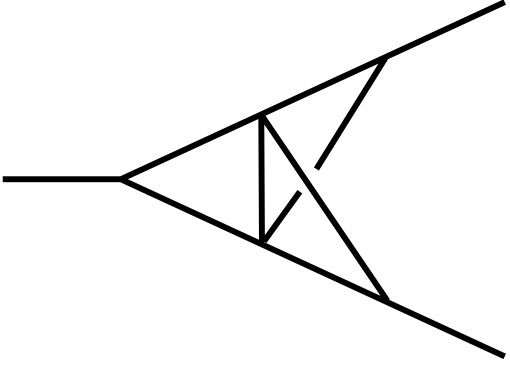}
        \caption*{$A_{7,4} \equiv A_{2,011110001101}$}
    \end{subfigure}

\vspace{0.5cm}

    \begin{subfigure}[b]{0.22\textwidth}
        \includegraphics[width=\textwidth]{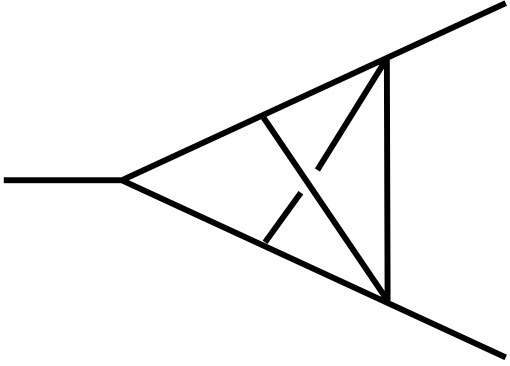}
        \caption*{$A_{7,5} \equiv A_{2,011011010101}$}
    \end{subfigure}
    ~~ 
    \begin{subfigure}[b]{0.22\textwidth}
        \includegraphics[width=\textwidth]{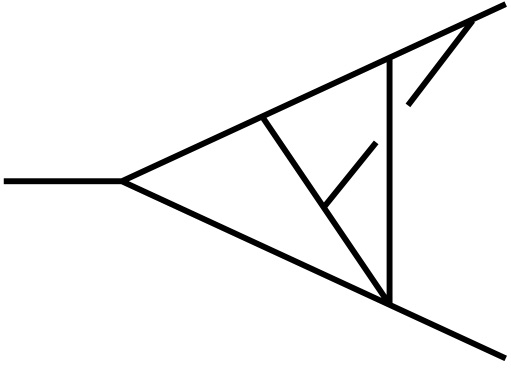}
        \caption*{$A_{8,1} \equiv A_{2,001111011101}$}
    \end{subfigure}

\vspace{0.5cm}

    \begin{subfigure}[b]{0.22\textwidth}
        \includegraphics[width=\textwidth]{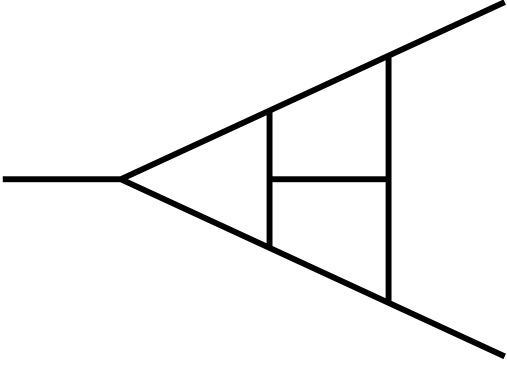}
        \caption*{$A_{9,1} \equiv A_{1,011111110110}$}
    \end{subfigure}
    ~~  
    \begin{subfigure}[b]{0.22\textwidth}
        \includegraphics[width=\textwidth]{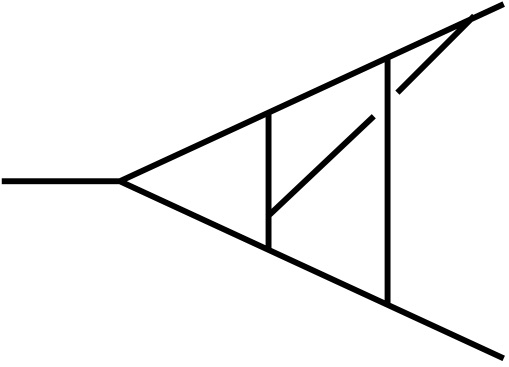}
        \caption*{$A_{9,2} \equiv A_{2,011111011101}$}
    \end{subfigure}
    ~~ 
    \begin{subfigure}[b]{0.22\textwidth}
        \includegraphics[width=\textwidth]{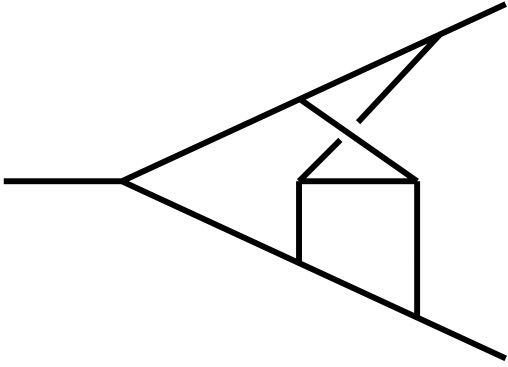}
        \caption*{$A_{9,4} \equiv A_{2,111011111100}$}
    \end{subfigure}
    \caption{Genuine three loop integrals with vertex function}
    \label{fig:A}
\end{figure}

We find that the form factors at three loop level can be expressed in terms of 22 master integrals.  Following the same notation as of \cite{Gehrmann:2010ue}, the master integrals can be distinguished into three topological types: genuine three loop integrals with vertex functions ($A_{t,i}$), three loop propagator integrals ($B_{t,i}$) and integrals which are product of one loop and two loop integrals ($C_{t,i}$).
Defining a generic three loop master integral as
\begin{align}
A_{i, m_{1}^i m_{2}^i \cdots m_{12}^i} = \int \frac{d^d k_1}{(2 \pi)^d} 
\int \frac{d^d k_2}{(2 \pi)^d} 
\int \frac{d^d k_3}{(2 \pi)^d} 
\frac{1}{\prod_{j} D_j^{m_j^i} } ,   \quad \quad \quad i=1,2,3
\end{align}
where $D_j$ is the $j^{\rm th}$ element of the set $A_i$,
we can identify the resulting master integrals in our computation with those given in \cite{Gehrmann:2010ue} and they
are listed\footnote{These figure have been taken from \cite{Gehrmann:2010ue}.}
in Fig.~\ref{fig:BnC} and Fig.~\ref{fig:A}.

The master integrals were computed in \cite{Baikov:2009bg, Gehrmann:2010ue} to relevant orders
in $\epsilon$ and we have used them to complete our computation of the form factors up to three loop
level.  The electronic version of the results of both quark and gluon 
form factors in terms of the master
integrals $A_{i,j},B_{i,j}$ and $C_{i,j}$ for arbitrary $d$ is attached with the arXiv version.
In the next section, we present the three loop results for both  the form factors expanded in powers of $\epsilon$ along with
already known one and two loop results.  

\subsection{Results}
In this section we present one, two and three loop quark and gluon form factors after expanding
in powers of $\ep$ to relevant order.  The one and two loop results completely agree with \cite{deFlorian:2013sza} and the three loop ones are new.
\begin{align}
\hat{\cal F}^{T,(1)}_{g} &=
       C_{A} \Bigg\{
          - \frac{8}{\epsilon^2}  
       + \frac{1}{\epsilon}   
            \frac{22}{3}
       - \frac{203}{18}
          + \zeta_2
       + \epsilon   \Bigg(
            \frac{2879}{216}
          - \frac{7}{3} \zeta_3
          - \frac{11}{12} \zeta_2
          \Bigg)
       + \epsilon^2   \Bigg(
          - \frac{37307}{2592}
          + \frac{77}{36} \zeta_3
\nonumber\\
&
          + \frac{203}{144} \zeta_2
          + \frac{47}{80} \zeta_2^2
          \Bigg)
       + \epsilon^3   \Bigg(
            \frac{465143}{31104}
          - \frac{31}{20} \zeta_5
          - \frac{1421}{432} \zeta_3
          - \frac{2879}{1728} \zeta_2
          + \frac{7}{24} \zeta_2 \zeta_3
          - \frac{517}{960} \zeta_2^2
          \Bigg)
\nonumber\\
&
       + \epsilon^4   \Bigg(
          - \frac{5695811}{373248}
          + \frac{341}{240} \zeta_5
          + \frac{20153}{5184} \zeta_3
          - \frac{49}{144} \zeta_3^2
          + \frac{37307}{20736} \zeta_2
          - \frac{77}{288} \zeta_2 \zeta_3
          + \frac{9541}{11520} \zeta_2^2
\nonumber\\
&
          + \frac{949}{4480} \zeta_2^3
          \Bigg)
         \Bigg\}
   + n_{f} \Bigg\{
          - \frac{4}{3} \frac{1}{\epsilon}
       + \frac{35}{18}
       + \epsilon   \Bigg(
          - \frac{497}{216}
          + \frac{1}{6} \zeta_2
          \Bigg)
       + \epsilon^2   \Bigg(
            \frac{6593}{2592}
          - \frac{7}{18} \zeta_3
          - \frac{35}{144} \zeta_2
          \Bigg)
\nonumber\\
&
       + \epsilon^3   \Bigg(
          - \frac{84797}{31104}
          + \frac{245}{432} \zeta_3
          + \frac{497}{1728} \zeta_2
          + \frac{47}{480} \zeta_2^2
          \Bigg)
       + \epsilon^4   \Bigg(
            \frac{1072433}{373248}
          - \frac{31}{120} \zeta_5
          - \frac{3479}{5184} \zeta_3
\nonumber\\
&
          - \frac{6593}{20736} \zeta_2
          + \frac{7}{144} \zeta_2 \zeta_3
          - \frac{329}{2304} \zeta_2^2
          \Bigg)
         \Bigg\} \,,
\nonumber\\
\hat{\cal F}^{T,(1)}_{q} &=
C_{F}  \Bigg\{ 
       - \frac{8}{\epsilon^2}  
       + \frac{6}{\epsilon}  
       - 10
          + \zeta_2
       + \epsilon   \Bigg(
            12
          - \frac{7}{3} \zeta_3
          - \frac{3}{4} \zeta_2
          \Bigg)
       + \epsilon^2   \Bigg(
          - 13
          + \frac{7}{4} \zeta_3
          + \frac{5}{4} \zeta_2
          + \frac{47}{80} \zeta_2^2
          \Bigg)
\nonumber\\
&
       + \epsilon^3   \Bigg(
            \frac{27}{2}
          - \frac{31}{20} \zeta_5
          - \frac{35}{12} \zeta_3
          - \frac{3}{2} \zeta_2
          + \frac{7}{24} \zeta_2 \zeta_3
          - \frac{141}{320} \zeta_2^2
          \Bigg)
       + \epsilon^4   \Bigg(
          - \frac{55}{4}
          + \frac{93}{80} \zeta_5
          + \frac{7}{2} \zeta_3
\nonumber\\
&
          - \frac{49}{144} \zeta_3^2
          + \frac{13}{8} \zeta_2
          - \frac{7}{32} \zeta_2 \zeta_3
          + \frac{47}{64} \zeta_2^2
          + \frac{949}{4480} \zeta_2^3
          \Bigg)
         \Bigg\} \,,
\end{align}

\begin{align}
\hat{\cal F}^{T,(2)}_{g} &=
       C_{A}^2  \Bigg\{ 
        \frac{32}{\epsilon^4}   
       - \frac{44}{\epsilon^3}  
       + \frac{1}{\epsilon^2}   \Bigg(
            \frac{226}{3}
          - 4 \zeta_2
          \Bigg)
       + \frac{1}{\epsilon}   \Bigg(
          - 81
          + \frac{50}{3} \zeta_3
          + \frac{11}{3} \zeta_2
          \Bigg)
       + \frac{5249}{108}
          - 11 \zeta_3
\nonumber\\
&
          - \frac{67}{18} \zeta_2
          - \frac{21}{5} \zeta_2^2
       + \epsilon   \Bigg(
            \frac{59009}{1296}
          - \frac{71}{10} \zeta_5
          + \frac{433}{18} \zeta_3
          - \frac{337}{108} \zeta_2
          - \frac{23}{6} \zeta_2 \zeta_3
          + \frac{99}{40} \zeta_2^2
          \Bigg)
\nonumber\\
&
       + \epsilon^2   \Bigg(
          - \frac{1233397}{5184}
          + \frac{759}{20} \zeta_5
          - \frac{8855}{216} \zeta_3
          + \frac{901}{36} \zeta_3^2
          + \frac{12551}{648} \zeta_2
          + \frac{77}{36} \zeta_2 \zeta_3
          - \frac{4843}{720} \zeta_2^2
\nonumber\\
&
          + \frac{2313}{280} \zeta_2^3
          \Bigg)
         \Bigg\}
   + C_{A} n_{f}  \Bigg\{ 
         \frac{8}{\epsilon^3}   
       + \frac{1}{\epsilon^2}   \Bigg(
          - \frac{40}{3}
          \Bigg)
       + \frac{1}{\epsilon}   \Bigg(
            \frac{41}{3}
          - \frac{2}{3} \zeta_2
          \Bigg)
       - \frac{605}{108}
          + 10 \zeta_3
          + \frac{5}{9} \zeta_2
\nonumber\\
&
       + \epsilon   \Bigg(
          - \frac{21557}{1296}
          - \frac{182}{9} \zeta_3
          + \frac{145}{108} \zeta_2
          - \frac{57}{20} \zeta_2^2
          \Bigg)
       + \epsilon^2   \Bigg(
            \frac{320813}{5184}
          + \frac{71}{10} \zeta_5
          + \frac{6407}{216} \zeta_3
          - \frac{3617}{648} \zeta_2
\nonumber\\
&
          - \frac{43}{18} \zeta_2 \zeta_3
          + \frac{1099}{180} \zeta_2^2
          \Bigg)
         \Bigg\}
   + C_{F} n_{f}  \Bigg\{ 
       - \frac{2}{\epsilon}
       + \frac{61}{6}
          - 8 \zeta_3
       + \epsilon   \Bigg(
          - \frac{2245}{72}
          + \frac{59}{3} \zeta_3
          + \frac{1}{2} \zeta_2
\nonumber\\
&
          + \frac{12}{5} \zeta_2^2
          \Bigg)
       + \epsilon^2   \Bigg(
            \frac{64177}{864}
          - 14 \zeta_5
          - \frac{335}{9} \zeta_3
          - \frac{83}{24} \zeta_2
          + 2 \zeta_2 \zeta_3
          - \frac{179}{30} \zeta_2^2
          \Bigg)
         \Bigg\} \,,
%
\\
\hat{\cal F}^{T,(2)}_{q} &=
       C_{F}^2  \Bigg\{ 
         \frac{32}{\epsilon^4}  
       - \frac{48}{\epsilon^3}  
       + \frac{1}{\epsilon^2}   \Bigg(
            98
          - 8 \zeta_2
          \Bigg)
       + \frac{1}{\epsilon}   \Bigg(
          - \frac{309}{2}
          + \frac{128}{3} \zeta_3
          \Bigg)
       + \frac{5317}{24}
          - 90 \zeta_3
          + \frac{41}{2} \zeta_2
\nonumber\\
&
          - 13 \zeta_2^2
       + \epsilon   \Bigg(
          - \frac{28127}{96}
          + \frac{92}{5} \zeta_5
          + \frac{1327}{6} \zeta_3
          - \frac{1495}{24} \zeta_2
          - \frac{56}{3} \zeta_2 \zeta_3
          + \frac{173}{6} \zeta_2^2
          \Bigg)
\nonumber\\
&
       + \epsilon^2   \Bigg(
            \frac{1244293}{3456}
          - \frac{311}{10} \zeta_5
          - \frac{34735}{72} \zeta_3
          + \frac{652}{9} \zeta_3^2
          + \frac{38543}{288} \zeta_2
          + \frac{193}{6} \zeta_2 \zeta_3
          - \frac{10085}{144} \zeta_2^2
\nonumber\\
&
          + \frac{223}{20} \zeta_2^3
          \Bigg)
         \Bigg\} 
   + C_{A} C_{F}  \Bigg\{
         \frac{44}{3} \frac{1}{\epsilon^3}  
       + \frac{1}{\epsilon^2}   \Bigg(
          - \frac{332}{9}
          + 4 \zeta_2
          \Bigg)
       + \frac{1}{\epsilon}   \Bigg(
            \frac{4921}{54}
          - 26 \zeta_3
          + \frac{11}{3} \zeta_2
          \Bigg)
\nonumber\\
&
       - \frac{120205}{648}
          + \frac{755}{9} \zeta_3
          - \frac{251}{9} \zeta_2
          + \frac{44}{5} \zeta_2^2
       + \epsilon   \Bigg(
            \frac{2562925}{7776}
          - \frac{51}{2} \zeta_5
          - \frac{5273}{27} \zeta_3
          + \frac{14761}{216} \zeta_2
\nonumber\\
&
          + \frac{89}{6} \zeta_2 \zeta_3
          - \frac{3299}{120} \zeta_2^2
          \Bigg)
       + \epsilon^2   \Bigg(
          - \frac{50471413}{93312}
          + \frac{3971}{60} \zeta_5
          + \frac{282817}{648} \zeta_3
          - \frac{569}{12} \zeta_3^2
          - \frac{351733}{2592} \zeta_2
\nonumber\\
&
          - \frac{1069}{36} \zeta_2 \zeta_3
          + \frac{7481}{120} \zeta_2^2
          - \frac{809}{280} \zeta_2^3
          \Bigg)
         \Bigg\}
   + C_{F} n_{f}  \Bigg\{  
       - \frac{8}{3} \frac{1}{\epsilon^3}  
       + \frac{56}{9} \frac{1}{\epsilon^2} 
       + \frac{1}{\epsilon}   \Bigg(
          - \frac{425}{27}
          - \frac{2}{3} \zeta_2
          \Bigg)
\nonumber\\
&
       + \frac{9989}{324}
          - \frac{26}{9} \zeta_3
          + \frac{38}{9} \zeta_2
       + \epsilon   \Bigg(
          - \frac{202253}{3888}
          + \frac{2}{27} \zeta_3
          - \frac{989}{108} \zeta_2
          + \frac{41}{60} \zeta_2^2
          \Bigg)
\nonumber\\
&
       + \epsilon^2   \Bigg(
            \frac{3788165}{46656}
          - \frac{121}{30} \zeta_5
          - \frac{935}{324} \zeta_3
          + \frac{22937}{1296} \zeta_2
          - \frac{13}{18} \zeta_2 \zeta_3
          + \frac{97}{180} \zeta_2^2
          \Bigg)
         \Bigg\}\,,
\end{align}

\begin{align}
\hat{\cal F}^{T,(3)}_{g} &=
     C_{A}^3  \Bigg\{ 
         \frac{1}{\epsilon^6}   \Bigg(
          - \frac{256}{3}
          \Bigg)
       + \frac{1}{\epsilon^5}   \Bigg(
            \frac{352}{3}
          \Bigg)
       + \frac{1}{\epsilon^4}   \Bigg(
          - \frac{14744}{81}
          \Bigg)
       + \frac{1}{\epsilon^3}   \Bigg(
            \frac{13126}{243}
          - \frac{176}{3} \zeta_3
          + \frac{484}{27} \zeta_2
          \Bigg)
\nonumber\\
&
       + \frac{1}{\epsilon^2}   \Bigg(
            \frac{149939}{486}
          - \frac{440}{27} \zeta_3
          - \frac{4321}{81} \zeta_2
          + \frac{494}{45} \zeta_2^2
          \Bigg)
       + \frac{1}{\epsilon}   \Bigg(
          - \frac{14639165}{17496}
          + \frac{1756}{15} \zeta_5
          - \frac{634}{9} \zeta_3
\nonumber\\
&
          + \frac{112633}{972} \zeta_2
          + \frac{170}{9} \zeta_2 \zeta_3
          + \frac{4213}{180} \zeta_2^2
          \Bigg)
       + \frac{1056263429}{1049760}
          + \frac{5014}{45} \zeta_5
          + \frac{539}{2430} \zeta_3
          - \frac{1766}{9} \zeta_3^2
\nonumber\\
&
          - \frac{1988293}{11664} \zeta_2
          - \frac{92}{9} \zeta_2 \zeta_3
          - \frac{64997}{2160} \zeta_2^2
          - \frac{22523}{270} \zeta_2^3
         \Bigg\}
   + C_{A}^2 n_{f}  \Bigg\{  
         \frac{1}{\epsilon^5}   \Bigg(
          - \frac{64}{3}
          \Bigg)
       + \frac{1}{\epsilon^4}   \Bigg(
            \frac{1840}{81}
          \Bigg)
\nonumber\\
&
       + \frac{1}{\epsilon^3}   \Bigg(
            \frac{5818}{243}
          - \frac{88}{27} \zeta_2
          \Bigg)
       + \frac{1}{\epsilon^2}   \Bigg(
          - \frac{56783}{486}
          - \frac{1456}{27} \zeta_3
          + \frac{892}{81} \zeta_2
          \Bigg)
       + \frac{1}{\epsilon}   \Bigg(
            \frac{3370273}{17496}
          + \frac{5831}{81} \zeta_3
\nonumber\\
&
          - \frac{26173}{972} \zeta_2
          + \frac{1153}{90} \zeta_2^2
          \Bigg)
       + \frac{5797271}{1049760}
          - \frac{11528}{45} \zeta_5
          + \frac{5401}{30} \zeta_3
          + \frac{489781}{11664} \zeta_2
          + \frac{2}{9} \zeta_2 \zeta_3
\nonumber\\
&
          - \frac{923}{72} \zeta_2^2
         \Bigg\}
   + C_{A} n_{f}^2  \Bigg\{ 
         \frac{1}{\epsilon^4}   \Bigg(
            \frac{160}{81}
          \Bigg)
       + \frac{1}{\epsilon^3}   \Bigg(
          - \frac{1340}{243}
          \Bigg)
       + \frac{1}{\epsilon^2}   \Bigg(
            7
          - \frac{4}{27} \zeta_2
          \Bigg)
       + \frac{1}{\epsilon}   \Bigg(
            \frac{45077}{8748}
\nonumber\\
&
          + \frac{940}{81} \zeta_3
          - \frac{5}{162} \zeta_2
          \Bigg)
       - \frac{32220173}{524880}
          - \frac{122141}{2430} \zeta_3
          + \frac{661}{216} \zeta_2
          - \frac{1777}{540} \zeta_2^2
         \Bigg\}
\nonumber\\
&
   + C_{A} C_{F} n_{f}  \Bigg\{ 
         \frac{1}{\epsilon^3}   \Bigg(
            \frac{128}{9}
          \Bigg)
       + \frac{1}{\epsilon^2}   \Bigg(
          - \frac{1712}{27}
          + \frac{512}{9} \zeta_3
          \Bigg)
       + \frac{1}{\epsilon}   \Bigg(
            \frac{11732}{81}
          - \frac{2360}{27} \zeta_3
          - \frac{8}{3} \zeta_2
\nonumber\\
&
          - \frac{256}{15} \zeta_2^2
          \Bigg)
       - \frac{152656}{1215}
          + \frac{1256}{9} \zeta_5
          - \frac{10754}{405} \zeta_3
          + \frac{34}{3} \zeta_2
          + \frac{132}{5} \zeta_2^2
         \Bigg\}
   + C_{F}^2 n_{f}  \Bigg\{ 
         \frac{1}{\epsilon}   \Bigg(
            \frac{2}{3}
          \Bigg)
\nonumber\\
&
       - \frac{241}{18}
          + 80 \zeta_5
          - \frac{148}{3} \zeta_3
         \Bigg\}
   + C_{F} n_{f}^2  \Bigg\{ 
         \frac{1}{\epsilon^2}   \Bigg(
          - \frac{16}{9}
          \Bigg)
       + \frac{1}{\epsilon}   \Bigg(
            \frac{388}{27}
          - \frac{32}{3} \zeta_3
          \Bigg)
       - \frac{5623}{81}
\nonumber\\
&
          + \frac{412}{9} \zeta_3
          + \frac{2}{3} \zeta_2
          + \frac{16}{5} \zeta_2^2
         \Bigg\} \,,
\\
\hat{\cal F}^{T,(3)}_{q} &=
      C_{F}^3  \Bigg\{ 
         \frac{1}{\epsilon^6}   \Bigg(
          - \frac{256}{3}
          \Bigg)
       + \frac{1}{\epsilon^5}   \Bigg(
            192
          \Bigg)
       + \frac{1}{\epsilon^4}   \Bigg(
          - 464
          + 32 \zeta_2
          \Bigg)
       + \frac{1}{\epsilon^3}   \Bigg(
            888
          - \frac{800}{3} \zeta_3
          + 24 \zeta_2
          \Bigg)
\nonumber\\
&
       + \frac{1}{\epsilon^2}   \Bigg(
          - \frac{4582}{3}
          + 808 \zeta_3
          - 258 \zeta_2
          + \frac{426}{5} \zeta_2^2
          \Bigg)
       + \frac{1}{\epsilon}   \Bigg(
            \frac{14375}{6}
          - \frac{1288}{5} \zeta_5
          - \frac{6854}{3} \zeta_3
\nonumber\\
&
          + \frac{2629}{3} \zeta_2
          + \frac{428}{3} \zeta_2 \zeta_3
          - \frac{7199}{30} \zeta_2^2
          \Bigg)
       - \frac{765629}{216}
          + \frac{12074}{15} \zeta_5
          + \frac{47557}{9} \zeta_3
          - \frac{1826}{3} \zeta_3^2
\nonumber\\
&
          - \frac{78665}{36} \zeta_2
          - \frac{361}{3} \zeta_2 \zeta_3
          + \frac{201691}{360} \zeta_2^2
          - \frac{9095}{252} \zeta_2^3
         \Bigg\}
   + C_{A}^2 C_{F}  \Bigg\{ 
         \frac{1}{\epsilon^4}   \Bigg(
          - \frac{3872}{81}
          \Bigg)
\nonumber\\
&
       + \frac{1}{\epsilon^3}   \Bigg(
            \frac{52168}{243}
          - \frac{704}{27} \zeta_2
          \Bigg)
       + \frac{1}{\epsilon^2}   \Bigg(
          - \frac{187292}{243}
          + \frac{6688}{27} \zeta_3
          - \frac{2212}{81} \zeta_2
          - \frac{352}{45} \zeta_2^2
          \Bigg)
\nonumber\\
&
       + \frac{1}{\epsilon}   \Bigg(
            \frac{4856336}{2187}
          + \frac{272}{3} \zeta_5
          - \frac{36884}{27} \zeta_3
          + \frac{120769}{243} \zeta_2
          + \frac{176}{9} \zeta_2 \zeta_3
          - \frac{1604}{15} \zeta_2^2
          \Bigg)
       - \frac{71947001}{13122}
\nonumber\\
&
          - \frac{2588}{9} \zeta_5
          + \frac{2464213}{486} \zeta_3
          - \frac{1136}{9} \zeta_3^2
          - \frac{1479931}{729} \zeta_2
          - \frac{926}{9} \zeta_2 \zeta_3
          + \frac{54071}{108} \zeta_2^2
          - \frac{6152}{189} \zeta_2^3
         \Bigg\}
\nonumber\\
&
   + C_{A} C_{F}^2  \Bigg\{ 
         \frac{1}{\epsilon^5}   \Bigg(
          - \frac{352}{3}
          \Bigg)
       + \frac{1}{\epsilon^4}   \Bigg(
            \frac{3448}{9}
          - 32 \zeta_2
          \Bigg)
       + \frac{1}{\epsilon^3}   \Bigg(
          - \frac{29620}{27}
          + 208 \zeta_3
          + \frac{28}{3} \zeta_2
          \Bigg)
\nonumber\\
&
       + \frac{1}{\epsilon^2}   \Bigg(
            \frac{207442}{81}
          - 1096 \zeta_3
          + \frac{2471}{9} \zeta_2
          - \frac{332}{5} \zeta_2^2
          \Bigg)
       + \frac{1}{\epsilon}   \Bigg(
          - \frac{2529065}{486}
          + 284 \zeta_5
          + \frac{10603}{3} \zeta_3
\nonumber\\
&
          - \frac{71101}{54} \zeta_2
          - \frac{430}{3} \zeta_2 \zeta_3
          + \frac{66091}{180} \zeta_2^2
          \Bigg)
       + \frac{56048957}{5832}
          - \frac{18274}{45} \zeta_5
          - \frac{185921}{18} \zeta_3
          + \frac{1616}{3} \zeta_3^2
\nonumber\\
&
          + \frac{1324001}{324} \zeta_2
          + \frac{1870}{9} \zeta_2 \zeta_3
          - \frac{2254603}{2160} \zeta_2^2
          - \frac{18619}{1260} \zeta_2^3
         \Bigg\}
   + C_{A} C_{F} n_{f}  \Bigg\{  
         \frac{1}{\epsilon^4}   \Bigg(
            \frac{1408}{81}
          \Bigg)
\nonumber\\
&
       + \frac{1}{\epsilon^3}   \Bigg(
          - \frac{18032}{243}
          + \frac{128}{27} \zeta_2
          \Bigg)
       + \frac{1}{\epsilon^2}   \Bigg(
            \frac{64220}{243}
          - \frac{1024}{27} \zeta_3
          + \frac{1264}{81} \zeta_2
          \Bigg)
       + \frac{1}{\epsilon}   \Bigg(
          - \frac{1613122}{2187}
\nonumber\\
&
          + \frac{19784}{81} \zeta_3
          - \frac{41062}{243} \zeta_2
          + \frac{88}{5} \zeta_2^2
          \Bigg)
       + \frac{11339972}{6561}
          - \frac{128}{3} \zeta_5
          - \frac{64730}{81} \zeta_3
          + \frac{916217}{1458} \zeta_2
\nonumber\\
&
          + \frac{392}{9} \zeta_2 \zeta_3
          - \frac{2161}{27} \zeta_2^2
         \Bigg\}
   + C_{F}^2 n_{f}  \Bigg\{  
         \frac{1}{\epsilon^5}   \Bigg(
            \frac{64}{3}
          \Bigg)
       + \frac{1}{\epsilon^4}   \Bigg(
          - \frac{592}{9}
          \Bigg)
       + \frac{1}{\epsilon^3}   \Bigg(
            \frac{5080}{27}
          + \frac{8}{3} \zeta_2
          \Bigg)
\nonumber\\
&
       + \frac{1}{\epsilon^2}   \Bigg(
          - \frac{34060}{81}
          + \frac{584}{9} \zeta_3
          - \frac{458}{9} \zeta_2
          \Bigg)
       + \frac{1}{\epsilon}   \Bigg(
            \frac{194287}{243}
          - \frac{5978}{27} \zeta_3
          + \frac{5651}{27} \zeta_2
          - \frac{337}{18} \zeta_2^2
          \Bigg)
\nonumber\\
&
       - \frac{3887467}{2916}
          + \frac{278}{45} \zeta_5
          + \frac{70499}{81} \zeta_3
          - \frac{99691}{162} \zeta_2
          - \frac{343}{9} \zeta_2 \zeta_3
          + \frac{49001}{1080} \zeta_2^2
         \Bigg\}
\nonumber\\
&
   + C_{F} n_{f}^2  \Bigg\{  
         \frac{1}{\epsilon^4}   \Bigg(
          - \frac{128}{81}
          \Bigg)
       + \frac{1}{\epsilon^3}   \Bigg(
            \frac{1504}{243}
          \Bigg)
       + \frac{1}{\epsilon^2}   \Bigg(
          - \frac{592}{27}
          - \frac{16}{9} \zeta_2
          \Bigg)
       + \frac{1}{\epsilon}   \Bigg(
            \frac{128312}{2187}
          - \frac{272}{81} \zeta_3
\nonumber\\
&
          + \frac{380}{27} \zeta_2
          \Bigg)
       - \frac{857536}{6561}
          - \frac{2852}{243} \zeta_3
          - \frac{1250}{27} \zeta_2
          - \frac{83}{135} \zeta_2^2
         \Bigg\} \,,
\end{align}
where $C_A = N$, $C_F = (N^2 -1)/{2 N}$ and $n_f$ is the number of active quark flavors. 
In the next section, we describe
how these form factors can be renormalized up to three loop level through coupling constant renormalization.
We then study the universal structure of the infrared poles in $\epsilon$ 
through Sudakov's KG equation 
up to three loop level.  It provides a crucial check for our new results on the form factors.

\section{Ultraviolet Renormalization}
\label{sec:uv}

In $\overline {MS}$ scheme, the renormalized coupling constant $a_s \equiv a_s (\mu_R^2)$ at the
renormalization scale $\mu_R$ is related to unrenormalized coupling constant $\hat{a}_s$ by 
\begin{align} \label{renas}
 \frac{\hat{a}_s}{\mu_0^{\epsilon}} S_{\epsilon} &= \frac{a_s}{\mu_R^{\epsilon}} {Z}(\mu_R^2)
\nonumber\\[1ex]
 &= \frac{a_s}{\mu_R^{\epsilon}} \left[   1 + a_s r_1 + a_s^2 r_2 + {\cal O}(a_s^3)  \right] 
\end{align}
where,
\begin{equation*}
 r_1 = \frac{2 \b0}{\ep} \;, \quad
 r_2 = \left( \frac{4\b0^2}{\ep^2} + \frac{\beta_1}{\ep} \right) \,,
\end{equation*}
where $\beta_i$ are the coefficients of QCD beta function :
\begin{equation}
 \b0 = \left(\frac{11}{3} C_A - \frac{2}{3} n_f\right) , \hspace{0.5cm}
 \beta_1 =  \left(\frac{34}{3} C_A^2 - \frac{10}{3} C_A n_f - 2 C_F n_f\right) \,.
\end{equation}
Using the Eq.~\ref{renas}, we now can express $M_{\rm I}$ (Eq.~\ref{eq:miexp}) in powers  
of renormalized $a_s$ with UV finite matrix elements $\rnM^{(i)}_{\rm I}$  
\begin{equation}
 M_{\rm I} = \Bigg( \rnM^{(0)}_{\rm I} + a_s \rnM^{(1)}_{\rm I} + a_s^{2} \rnM^{(2)}_{\rm I} + a_s^{3} \rnM^{(3)}_{\rm I} + {\cal O}({a}_s^4) \Bigg)
\end{equation}
where,
\begin{align}
 \rnM^{(0)}_{\rm I} &= \unM_{\rm I}^{(0)} \,,
 \nonumber\\
 \rnM^{(1)}_{\rm I} &= \left( \frac{Q^2}{\mu_R^2} \right)^{\frac{\ep}{2}} \unM_{\rm I}^{(1)} \,,
 \nonumber\\
 \rnM^{(2)}_{\rm I} &= \left( \frac{Q^2}{\mu_R^2} \right)^{\ep} \unM_{\rm I}^{(2)} 
             + r_1 \left( \frac{Q^2}{\mu_R^2} \right)^{\frac{\ep}{2}} \unM_{\rm I}^{(1)} \,,
 \nonumber\\
 \rnM^{(3)}_{\rm I} &= \left( \frac{Q^2}{\mu_R^2} \right)^{\frac{3\ep}{2}} \unM_{\rm I}^{(3)}
              + 2 r_1 \left( \frac{Q^2}{\mu_R^2} \right)^{\ep} \unM_{\rm I}^{(2)}
              + r_2 \left( \frac{Q^2}{\mu_R^2} \right)^{\frac{\ep}{2}} \unM_{\rm I}^{(1)} \,.
\end{align}
Using above equations, we can obtain 
the renormalized form factors ${\cal{F}}^{\rm T}_{\rm I}$ in terms of $a_s$.

\section{Infrared Singularities and Universal Pole Structure}
\label{sec:ir}

The results on multiparton amplitudes beyond leading order in perturbative
QCD have not only played an important role in understanding the 
infrared structure of the theory but also allowed us to successfully carry out
various resummation programs for physical observables in the kinematic regions 
where the fixed order perturbation theory breaks down.   
The most important one along this line was the very successful proposal by Catani \cite{Catani:1998bh}
(see also \cite{Sterman:2002qn}) on one and two loop QCD amplitudes using the universal subtraction
operators.  The generalization of this proposal was achieved by Becher and Neubert \cite{Becher:2009cu}
and also by Gardi and Magnea \cite{Gardi:2009qi} beyond two loops.  In \cite{Ravindran:2004mb}, for the first time, the structure of
single pole term in both quark and gluon form factors at two loop level was unraveled. 
It was shown explicitly that the single pole can be written as a linear combination of UV, 
collinear and soft anomalous dimensions.  The fact that this feature continues to hold even 
at three loop level for the same form factors was observed in \cite{Moch:2005tm}.  
The structure of the single pole term for the multiparton amplitudes was studied in detail 
in \cite{Aybat:2006wq, Aybat:2006mz}.  

The form factors $\hat{\cal F}^{T}_{\rm I} (\hat{a}_s, Q^2, \mu^2, \epsilon)$ satisfy the $KG$-differential equation which follows from the factorization property, gauge and renormalization group invariances \cite{Sudakov:1954sw, Mueller:1979ih, Collins:1980ih, Sen:1981sd} 
\begin{equation}
 \label{eq:KG}
 Q^2 \frac{d}{dQ^2} \ln \hat{\cal F}^{T}_{\rm I} (\hat{a}_s, Q^2, \mu^2, \epsilon)  = \frac{1}{2} \left[ K^{\rm T,I} (\hat{a}_s, \frac{\mu_R^2}{\mu^2}, \epsilon ) + G^{\rm T,I} (\hat{a}_s, \frac{Q^2}{\mu_R^2}, \frac{\mu_R^2}{\mu^2}, \epsilon ) \right]
\end{equation}
where, all the poles in dimensional regularization parameter $\ep$ are contained in $K^{\rm T,I}$ 
which is also taken to be $Q^2$ independent and the finite terms as $\ep \rightarrow 0$ are 
encapsulated in $G^{\rm T,I}$. Renormalization group invariance of the form factor implies
\begin{equation}
\label{eq:KIA}
\mu_R^2 \frac{d}{d\mu_R^2} K^{\rm T,I} (\hat{a}_s, \frac{\mu_R^2}{\mu^2}, \epsilon ) = - \mu_R^2 \frac{d}{d\mu_R^2} G^{\rm T,I} (\hat{a}_s, \frac{Q^2}{\mu_R^2}, \frac{\mu_R^2}{\mu^2}, \epsilon ) 
= - \sum_{i=1}^{\infty}  a_s^i (\mu_R^2) A^{\rm T,I}_i \,,
\end{equation}
where, $A^{\rm T,I}$ are the cusp anomalous dimensions. 
Since $K^{I}$ in Eq.~\ref{eq:KIA} contains only poles in $\ep$ with no $Q^2$ dependence, it can be
easily solved in powers of bare strong coupling constant $\hat{a}_s$.  Expressing 
\begin{align} \label{eq:ki}
K^{\rm T,I} \left( {\hat a}_{s}, \frac{\mu_{R}^{2}}{\mu^{2}}, \ep \right) = \sum_{i=1}^{\infty} {\hat a}_{s}^{i} \left(\frac{\mu_{R}^{2}}{\mu^{2}}\right)^{i\frac{\ep}{2}} S_{\ep}^{i} K^{{\rm T,I};(i)}(\ep)\, ,
\end{align}
we find that the constants $K^{{\rm T,I};(i)}(\ep)$ consist of simple poles in $\ep$ with the coefficients containing $A_{i}^{\rm T,I}$' and $\beta_{i}$'s. These can readily be found in \cite{Ravindran:2005vv, Ravindran:2006cg}.

The renormalization group equation of 
$G^{\rm T,I}(\hat{a}_s, \frac{Q^2}{\mu_R^2}, \frac{\mu_R^2}{\mu^2}, \epsilon )$ 
can be solved by imposing the boundary condition at $\mu_R^2=Q^2$.   Hence the solution can
be expressed in terms of the boundary function $G^{\rm T,I} (a_s(Q^2), 1, \epsilon)$ and
the term that contains full $\mu_{R}^{2}$ dependence :  

\begin{align} \label{eq:gi}
G^{\rm T,I}(\hat{a}_s, \frac{Q^2}{\mu_R^2}, \frac{\mu_R^2}{\mu^2}, \epsilon )
= G^{\rm T,I} (a_s(Q^2), 1, \epsilon) + \int_{\frac{Q^2}{\mu_R^2}}^1 \frac{dx}{x} A^{\rm T,I} (a_s (x \mu_R^2) )
\end{align}
The $\mu_R^2$ independent part of the solution can be expanded in powers of $a_{s}$ as 
\begin{align}
G^{\rm T,I} (a_s(Q^2), 1, \epsilon) = \sum_{i=1}^{\infty} a_s^i(Q^2) G^{\rm T,I}_i(\epsilon)\, .
\end{align}
Substituting Eq.~\ref{eq:ki} and Eq.~\ref{eq:gi} in Eq.~\ref{eq:KG}, and integrating over $Q^2$, we obtain the form factor
in powers of strong coupling constant:   
\begin{align}
\label{eq:lnFSoln}
\ln \hat{\cal F}^T_{\rm I} (\hat{a}_s, Q^2, \mu^2, \epsilon) = \sum_{i=1}^{\infty} {\hat a}_{s}^{i} \left( \frac{Q^{2}}{\mu^{2}}\right)^{i \frac{\ep}{2}} S_{\ep}^{i} \hat {\cal L}_{{\cal F}^T}^{{\rm I},(i)}(\ep)
\end{align}
with
\begin{align}
\label{eq:lnFitoCalLF}
\hat {\cal L}_{{\cal F}^T}^{{\rm I},(1)}(\ep) &= { \frac{1}{\ep^2} } \Bigg\{-2 A_1^{\rm T,I}\Bigg\}
              + { \frac{1}{\ep} } \Bigg\{G_1^{\rm T,I} (\ep)\Bigg\}\, ,
\nonumber\\
\hat {\cal L}_{{\cal F}^T}^{{\rm I},(2)}(\ep) &= { \frac{1}{\ep^3} } \Bigg\{\beta_0 A_1^{\rm T,I} \Bigg\}
                  + { \frac{1}{\ep^2} } \Bigg\{- { \frac{1}{2} } A_2^{\rm T,I}
                  - \beta_0  G_1^{\rm T,I} (\ep)\Bigg\}
                  + { \frac{1}{\ep} } \Bigg\{ { \frac{1}{2} } G_2^{\rm T,I} (\ep)\Bigg\}\, ,
\nonumber\\
\hat {\cal L}_{{\cal F}^T}^{{\rm I},(3)}(\ep) &= { \frac{1}{\ep^4} } \Bigg\{- { \frac{8}{9} } \beta_0^2 A_1^{\rm T,I} \Bigg\}
                  + { \frac{1}{\ep^3} } \Bigg\{ { \frac{2}{9} } \beta_1 A_1^{\rm T,I}
                    + { \frac{8}{9} } \beta_0 A_2^{\rm T,I} + { \frac{4}{3} }
                     \beta_0^2 G_1^{\rm T,I} (\ep)\Bigg\}
\nonumber\\
&                  + { \frac{1}{\ep^2} } \Bigg\{- { \frac{2}{9} } A_3^{\rm T,I}
                   - { \frac{1}{3} } \beta_1 G_1^{\rm T,I} (\ep)
                   - { \frac{4}{3} } \beta_0 G_2^{\rm T,I} (\ep)\Bigg\}
                   + { \frac{1}{\ep} } \Bigg\{ { \frac{1}{3} } G_3^{\rm T,I} (\ep)\Bigg\}\, .
\end{align}
It is now straightforward to extract the 
cusp anomalous dimensions by comparing Eq.~\ref{eq:lnFitoCalLF} with the form factors presented
in the previous section. We find

\begin{align}
 A^{\rm T, I}_1 &= {{C_I}} \Big\{4\Big\} \,, 
\nonumber\\
 A^{\rm T, I}_2 &= {{C_{\rm I} C_A}} \left\{ \frac{268}{9} - 8 \zeta_2 \right\} 
                 + {{C_I n_f}} \left\{ -\frac{40}{9} \right\} \,,
\nonumber\\
 A^{\rm T, I}_3 &= {{C_{\rm I} C_A^2}} \left\{ \frac{490}{3} 
- \frac{1072 \zeta_2 }{9} 
+ \frac{88 \zeta_3}{3} 
+ \frac{176 \zeta_2^2}{5} \right\}
+ {{C_{\rm I} C_F n_f}} \left\{ - \frac{110}{3} + 32 \zeta_3 \right\}
\nonumber\\ \nonumber
& + {{C_{\rm I} C_A n_f}} \left\{ - \frac{836}{27} 
+ \frac{160 \zeta_2}{9} 
- \frac{112 \zeta_3}{3} \right\}
+ {{C_{\rm I} n_f^2}} \left\{ - \frac{16}{27} \right\} \,.
\end{align}
where $C_{\rm I} = C_F$ for ${\rm I} = q$ and $C_{\rm I} = C_A$ for ${\rm I} = g$.
We find that they not only satisfy the property of maximally non-abelian but also
coincide with those that appear in the quark and gluon form factors which are 
available up to three-loop level in the literature \cite{Moch:2004pa, Vogt:2004mw}. 
\begin{align}
\label{eq:AgAq}
A_{i}^{{\rm T},g} = \frac{C_{A}}{C_{F}} A_{i}^{{\rm T},q} \, \quad \quad {\rm and} \quad \quad 
A_{i}^{\rm T,I} =  A_{i}^{\rm I}  \quad \quad \quad {\rm I}=q,g\, .
\end{align}
%
%
%
%
Following \cite{Moch:2005tm} and \cite{Ravindran:2004mb}, we can parametrize $G_i^{\rm T,I}(\ep)$ as follows:
\begin{align}
\label{eq:GIi}
 G^{\rm T,I}_i (\ep) = 2 \left(B^{\rm T,I}_i - \gamma_{i-1}^{\rm T,I}\right) 
                     + f^{\rm T,I}_i + C^{\rm T,I}_i + \sum_{k=1}^{\infty} \epsilon^k g_i^{{\rm T,I}; (k)} \, ,
\end{align}
where
the constants $C_{i}^{\rm T,I}$ are given by \cite{Ravindran:2006cg}
\begin{align}
\label{eq:Cg}
C_{1}^{\rm T,I} &= 0\, ,
\nonumber\\
C_{2}^{\rm T,I} &= - 2 \beta_{0} g_{1}^{{\rm T,I};(1)}\, ,
\nonumber\\
C_{3}^{\rm T,I} &= - 2 \beta_{1} g_{1}^{{\rm T,I};(1)} - 2 \beta_{0} \left(g_{2}^{{\rm T,I};(1)} + 2 \beta_{0} g_{1}^{{\rm T,I};(2)}\right)\, .
\end{align}
Using the above decomposition, 
we can extract $B_i^{\rm T,I}$ and $f_i^{\rm T,I}$
from the form factors computed up to three loop level.  They are found to be 
\begin{align}
B^{{\rm T},g}_1 &= {{C_A}} \left\{\frac{11}{3}\right\} - {{n_f}} \left\{\frac{2}{3}\right\}\, ,
\nonumber\\
B^{{\rm T},g}_2 &= {{C_A^2}} \left\{\frac{32}{3} + 12 \zeta_3\right\} - {{n_f C_A}} \left\{\frac{8}{3} \right\} - {{n_f C_F}} \Big\{2 \Big\}\, ,
\nonumber\\
B^{{\rm T},g}_3 &= {{C_A C_F n_f}} \left\{-\frac{241}{18}\right\}
        + {{C_A n_f^2}} \left\{\frac{29}{18}\right\}
        - {{C_A^2 n_f}} \left\{\frac{233}{18} + \frac{8}{3} \zeta_2+ \frac{4}{3} \zeta_2^2 + \frac{80}{3} \zeta_3\right\}
\nonumber\\       
&+ {{C_A^3}} \left\{\frac{79}{2} - 16 \zeta_2 \zeta_3 + \frac{8}{3} \zeta_2 + \frac{22}{3} \zeta_2^2
        + \frac{536}{3} \zeta_3 - 80 \zeta_5\right\}
        + {{C_F n_f^2}} \left\{\frac{11}{9}\right\}
        + {{C_F^2 n_f}} \Big\{{1}\Big\}\, ,
\nonumber\\
B^{{\rm T},q}_1 &= {{C_F}} \Big\{{1}\Big\}\,,
\nonumber \\
B^{{\rm T},q}_2 &=   {{C_F^2}} \Bigg\{ \frac{3}{2} - 12 \zeta_2 + 24 \zeta_3 \Bigg\}
              + {{C_A C_F}} \Bigg\{ \frac{17}{34} + \frac{88}{6} \zeta_2 - 12 \zeta_3 \Bigg\}
                + {{n_f C_F}}T_{F} \Bigg\{ - \frac{2}{3} - \frac{16}{3} \zeta_2 \Bigg\}\,,
\nonumber \\
 B^{{\rm T},q}_3 &=  {{C_A^2 {C_F}}} \Bigg\{ - 2 {\zeta_2}^2 + \frac{4496}{27} {\zeta_2}
          - \frac{1552}{9} {\zeta_3} + 40 {\zeta_5} - \frac{1657}{36} \Bigg\}
          + {{{C_A} C_F^2}} \Bigg\{ -\frac{988}{15} {\zeta_2}^2 + 16 {\zeta_2} {\zeta_3} 
\nonumber\\
&
          - \frac{410}{3} {\zeta_2} +\frac{844}{3} {\zeta_3}
          + 120 {\zeta_5} + \frac{151}{4} \Bigg\}
+ {{{C_A} {C_F} {n_f}}} \Bigg\{ \frac{4}{5} {\zeta_2}^2 - \frac{1336}{27} {\zeta_2}
          + \frac{200}{9} {\zeta_3} + 20 \Bigg\}
\nonumber\\
&
+ {{C_F^3}} \Bigg\{ \frac{288}{5} {\zeta_2}^2 - 32 {\zeta_2} {\zeta_3}
          + 18 {\zeta_2} + 68 {\zeta_3} - 240 {\zeta_5} + \frac{29}{2} \Bigg\}
\nonumber\\
&
+ {{C_F^2 {n_f}}} \Bigg\{ \frac{232}{15} {\zeta_2}^2 + \frac{20}{3} {\zeta_2}
          -\frac{136}{3} {\zeta_3} - 23 \Bigg\}
+ {{{C_F} n_f^2}} \Bigg\{ \frac{80}{27} {\zeta_2} - \frac{16}{9} {\zeta_3} - \frac{17}{9} \Bigg\} \, .
\end{align}
We find that the above $B^{\rm T, I}_i$ are identical to the ones that appear in quark and gluon
form factors of \cite{Moch:2005tm}:
\begin{eqnarray}
B^{\rm T,I}_i= B^{{\rm I}}_i\,, \quad \quad \quad I=q,g,~ i=1,2,3
\end{eqnarray}
and  
\begin{align}
\label{eq:fg}
 f_1^{\rm T, I} &= 0 \,,
\nonumber \\
 f_2^{\rm T, I} &= {{C_{\rm I} C_A}} \left\{ -\frac{22}{3} {\zeta_2} - 28 {\zeta_3} + \frac{808}{27} \right\}
        + {{C_{\rm I} n_f}} \left\{ \frac{4}{3} {\zeta_2} - \frac{112}{27} \right\} \,,
\nonumber \\
 f_3^{\rm T, I} &= {{C_{\rm I} C_A^2}} \left\{ \frac{352}{5} {\zeta_2}^2 + \frac{176}{3} {\zeta_2} {\zeta_3}
        - \frac{12650}{81} {\zeta_2} - \frac{1316}{3} {\zeta_3} + 192 {\zeta_5}
        + \frac{136781}{729}\right\}
\nonumber \\
&
        + {{{C_{\rm I} C_A} {n_f}}} \left\{ - \frac{96}{5} {\zeta_2}^2 
        + \frac{2828}{81} {\zeta_2}
        + \frac{728}{27} {\zeta_3} - \frac{11842}{729} \right\} 
\nonumber \\ 
&
        + {{C_{\rm I} {C_F} {n_f}}} \left\{ \frac{32}{5} {\zeta_2}^2 + 4 {\zeta_2} 
        + \frac{304}{9} {\zeta_3} - \frac{1711}{27} \right\}
\nonumber \\ 
&
        + {{C_{\rm I} {n_f}^2}} \left\{ - \frac{40}{27} {\zeta_2} + \frac{112}{27} {\zeta_3}
        - \frac{2080}{729} \right\} \, .
\end{align}
Similar to cusp anomalous dimensions, we find that $f^{\rm T,I}_i$ satisfy 
the property of maximally non-abelian and in addition
they coincide with those that appear in the quark and gluon form factors which are
available up to three-loop level in the literature \cite{Ravindran:2004mb},
\begin{align}
\label{eq:fgfq}
f_{i}^{{\rm T},g} = \frac{C_{A}}{C_{F}} f_{i}^{{\rm T},q} \, \quad \quad {\rm and} \quad \quad 
f_{i}^{\rm T,I} =  f_{i}^{\rm I}  \quad \quad \quad {\rm I}=q,g\,  \quad \quad i=1,2,3.
\end{align}
The UV anomalous dimensions are found to be identically zero due the conservation
of QCD energy momentum tensor, i.e.,
\begin{align}
\gamma_{i}^{\rm T, I} = 0.
\end{align}
The universal behavior of infrared poles in terms of the cusp ($A^{\rm I}$), collinear ($B^{\rm I}$) and soft ($f^{\rm I}$) anomalous
dimensions provides a crucial check on our computation.  
The remaining terms namely $g_{i}^{{\rm T,I};(k)}$'s in Eq.~\ref{eq:GIi} can be extracted from the form factors and they are
listed below:

\begin{align}
 g_1^{{\rm T,g}; (1)} &=
       C_{A}   \Bigg(
          - \frac{203}{18}
          + \zeta_2
          \Bigg)
       + n_{f}   \Bigg(
            \frac{35}{18}
          \Bigg)\,,
\nonumber\\
 g_1^{{\rm T,g}; (2)} &= 
        C_{A}   \Bigg(
            \frac{2879}{216}
          - \frac{7}{3} \zeta_3
          - \frac{11}{12} \zeta_2
          \Bigg)
       + n_{f}   \Bigg(
          - \frac{497}{216}
          + \frac{1}{6} \zeta_2
          \Bigg)\,,
\nonumber\\
 g_1^{{\rm T,g}; (3)} &=
         C_{A}   \Bigg(
          - \frac{37307}{2592}
          + \frac{77}{36} \zeta_3
          + \frac{203}{144} \zeta_2
          + \frac{47}{80} \zeta_2^2
          \Bigg)
       + n_{f}   \Bigg(
            \frac{6593}{2592}
          - \frac{7}{18} \zeta_3
          - \frac{35}{144} \zeta_2
          \Bigg)\,,
\nonumber\\
g_2^{{\rm T,g}; (1)} &=
         C_{A}^2   \Bigg(
          - \frac{19333}{54}
          + \frac{88}{3} \zeta_3
          + \frac{799}{18} \zeta_2
          \Bigg)
       + C_{A} n_{f}   \Bigg(
            \frac{34991}{324}
          + \frac{32}{3} \zeta_3
          - \frac{82}{9} \zeta_2
          \Bigg)
\nonumber\\
&
       + C_{F} n_{f}   \Bigg(
            \frac{61}{3}
          - 16 \zeta_3
          \Bigg)
       + n_{f}^2   \Bigg(
          - \frac{2219}{324}
          + \frac{2}{9} \zeta_2
          \Bigg)\,,
\nonumber\\
g_2^{{\rm T,g}; (2)} &=
         C_{A}^2   \Bigg(
            \frac{2863591}{3888}
          - 39 \zeta_5
          - \frac{437}{6} \zeta_3
          - \frac{6521}{72} \zeta_2
          + \frac{5}{3} \zeta_2 \zeta_3
          - \frac{737}{120} \zeta_2^2
          \Bigg)
       + C_{A} n_{f}   \Bigg(
          - \frac{849385}{3888}
\nonumber\\
&
          - \frac{448}{27} \zeta_3
          + \frac{183}{8} \zeta_2
          - \frac{221}{60} \zeta_2^2
          \Bigg)
       + C_{F} n_{f}   \Bigg(
          - \frac{2245}{36}
          + \frac{118}{3} \zeta_3
          + \zeta_2
          + \frac{24}{5} \zeta_2^2
          \Bigg)
\nonumber\\
&
       + n_{f}^2   \Bigg(
            \frac{1999}{162}
          - \frac{14}{27} \zeta_3
          - \frac{35}{36} \zeta_2
          \Bigg)\,,
\end{align}

\begin{align}
%
g_1^{{\rm T,q}; (1)} &=
         C_{F}   \Bigg(
          - 10
          + \zeta_2
          \Bigg)\,,
\nonumber\\
 g_1^{{\rm T,q}; (2)} &=  
         C_{F}   \Bigg(
            12
          - \frac{7}{3} \zeta_3
          - \frac{3}{4} \zeta_2
          \Bigg)\,,
\nonumber\\
 g_1^{{\rm T,q}; (3)} &=  
         C_{F}   \Bigg(
          - 13
          + \frac{7}{4} \zeta_3
          + \frac{5}{4} \zeta_2
          + \frac{47}{80} \zeta_2^2
          \Bigg)\,,
\nonumber\\
 g_2^{{\rm T,q}; (1)} &= 
         C_{F}^2   \Bigg(
          - \frac{107}{12}
          - 124 \zeta_3
          + 90 \zeta_2
          - \frac{88}{5} \zeta_2^2
          \Bigg)
       + C_{A} C_{F}   \Bigg(
          - \frac{91693}{324}
          + \frac{452}{3} \zeta_3
          - \frac{1103}{18} \zeta_2
\nonumber\\
&
          + \frac{88}{5} \zeta_2^2
          \Bigg)
       + C_{F} n_{f}   \Bigg(
            \frac{7397}{162}
          - \frac{8}{3} \zeta_3
          + \frac{85}{9} \zeta_2
          \Bigg)\,,
\nonumber\\
  g_2^{{\rm T,q}; (2)} &= 
         C_{F}^2   \Bigg(
            \frac{1249}{48}
          + 12 \zeta_5
          + 328 \zeta_3
          - \frac{2431}{12} \zeta_2
          - 28 \zeta_2 \zeta_3
          + \frac{676}{15} \zeta_2^2
          \Bigg)
\nonumber\\
&
       + C_{A} C_{F}   \Bigg(
            \frac{2192269}{3888}
          - 51 \zeta_5
          - \frac{20399}{54} \zeta_3
          + \frac{15751}{108} \zeta_2
          + \frac{89}{3} \zeta_2 \zeta_3
          - \frac{2027}{40} \zeta_2^2
          \Bigg)
\nonumber\\
&
       + C_{F} n_{f}   \Bigg(
          - \frac{168557}{1944}
          - \frac{59}{27} \zeta_3
          - \frac{1079}{54} \zeta_2
          + \frac{7}{12} \zeta_2^2
          \Bigg)\,.
\end{align}

\section{Conclusions}
\label{sec:conclu}

We have presented both quark-antiquark and gluon-gluon form factors of the spin-2 fields
that couple to fields of SU(N) gauge theory with $n_f$ light flavors.  We have used
state-of-the-art methods to perform this computation efficiently as the number of
Feynman diagrams involved is quite large compared to other known form factors.
We have used IBP and LI identities to express the form factors
in terms of 22 master integrals.  We have presented the form factors in terms
of these master integrals for arbitrary $d$ as well as in powers of $\ep = d-4$ to appropriate
order, thanks to the availability of the master integrals to relevant orders in $\ep$
for further study.  These form factors are important components to the
scattering cross sections involving spin-2 fields beyond leading order
in QCD.   We have shown that these form factors do satisfy Sudakov integro-differential equation
and hence exhibit identical infrared structure of other form factors such as those appearing
in electroweak vector boson and Higgs productions up to three loop level.  We have also
shown these factors do not require overall renormalization due to the conservation
property of the energy momentum tensor.  Our results will be useful in improving the perturbative predictions
of spin-2 resonance production beyond NNLO level at the LHC where searches for
such particles are already underway with the upgraded energy and luminosity.

\section*{Acknowledgement}
We sincerely thank T. Gehrmann for constant encouragement as well as for providing us
all the master integrals for the present computation.  We would like to thank
R. Lee for his timely help with LiteRed.  We thank Hasegawa, Kumar, Mahakhud, Mandal,
Mazzitelli and Tancredi for discussions at various stages of this computation.  We acknowledge
IMSc computer facility for their support.

\end{document}